%% file: main_v19.tex
\documentclass[preprint,12pt]{elsarticle}

\usepackage{graphicx}

\usepackage{amssymb}

\journal{Nucl. Instr. Meth. A}

\begin{document}

\begin{frontmatter}



\title{XMASS detector}


\input{tech-auth.tex}

\begin{abstract}
The XMASS project aims to detect dark matter, 
pp and ${}^{7}$Be solar neutrinos, and neutrinoless double beta decay 
using ultra pure liquid xenon. 
The first phase of the XMASS experiment searches for dark matter.
In this paper, we describe the XMASS detector in detail, 
including its configuration, data acquisition equipment and calibration system.
\end{abstract}

\begin{keyword}

Revision 5
\end{keyword}

\end{frontmatter}


\section{Introduction}
XMASS is a project aimed at detecting dark matter, 
pp and ${}^{7}$Be solar neutrinos and neutrinoless double beta decay 
using ultra-pure liquid xenon. The original idea is presented in Ref.\ \cite{org-xmass}.
The first phase of the XMASS experiment concentrates on 
dark matter detection.
Various astronomical observations \cite{dmev} provide strong evidence 
for a large amount of dark matter in the universe. 
However, its nature is still unknown. 
Weakly interacting massive particles (WIMPs) are dark matter candidates
motivated by theories beyond the Standard Model \cite{wimpmodel}.
This project searches for nuclear recoils in liquid xenon 
caused by WIMPs.
The advantages of using liquid xenon as the target material are: 
First, liquid xenon yields a large amount 
of scintillation light, comparable to the yield of a NaI(Tl) 
scintillator. 
Second, xenon has a high atomic number $(Z=54)$, and liquid
xenon has a high density ($\sim$${\rm 2.9\,\rm g/cm^3}$).
Thus, the target volume can be small,
and external background (BG) gamma rays can be 
absorbed within a short distance from the detector wall. 
On the other hand, dark matter particles interact uniformly in the detector.
By extracting events that occur only in the interior volume of the detector, 
a sensitive search for dark matter particles can be conducted.

The XMASS detector is located at the Kamioka Observatory of the Institute
for Cosmic Ray Research, the University of Tokyo, Japan. 
It is in the Kamioka mine ${\rm 1,000 \, m}$ underneath the
top of Mt. Ikenoyama (i.e. ${\rm 2,700 \, m}$ water equivalent underground).
The detector has two components, the inner and outer detectors
(ID and OD, respectively).
The ID is equipped with 642 inward-facing photomultiplier tubes (PMTs)
in an approximate spherical shape in a copper vessel filled with pure liquid 
xenon.
The amount of liquid xenon in the sensitive region is 835 kg.
The ID is installed at the centre of the OD, which is
a cylindrical water tank with 72 twenty-inch PMTs.
The OD is used as an active shield for cosmic ray muons 
and a passive shield for low-energy gamma rays and neutrons.
Construction of the detector started in April 2007 and
was completed in September 2010.
Commissioning runs were conducted from October 2010 to June 2012.

In this paper, we describe the detector in detail, including the active water 
shield, ID structure, distillation system, cryogenic system, low-BG PMTs, 
data acquisition equipment, and calibration system.

\section{ Water shield }

\begin{figure}
\begin{center}
\includegraphics[width=11.0cm]{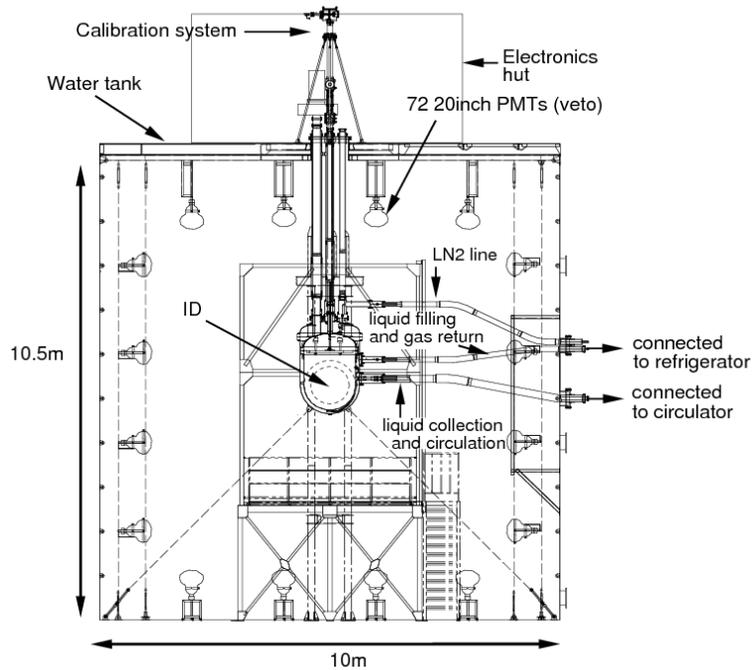}
\caption{OD consisting of water tank and 20-in. PMTs. The ID was installed at the centre of the water tank. Electronics hut on top of the tank records signals from PMTs in the ID and OD. A calibration system is also installed on top of the tank. A refrigerator that liquefies xenon gas and holds liquid xenon is located in the side of tank.}
\label{fig:fig_water}
\end{center}
\end{figure}

A water shield is more suitable than a conventional lead and copper shield 
because it produces fewer neutrons from 
cosmic rays and provides a greater reduction in incoming fast neutrons.
A cylindrical water tank ${\rm 10.5 \, m}$ in height and 
${\rm 10 \, m}$ in diameter is used for the radiation shield, and
72 twenty-inch PMTs (HAMAMATSU R3600) on the inner surface of the tank 
are used to actively reject cosmic rays (${\rm Fig. \, \ref{fig:fig_water}}$).
The ID was installed at the centre of the water tank, 
so the water shield is more than 4m thick in all directions.
The performance of the shield for gamma rays and neutrons 
from rock outside the tank 
was evaluated by a simulation study.
The study showed that ${\rm 2 \, m}$ of water shield
reduce the event rate from external fast neutrons as well as gamma rays
to levels that are insignificant compared to the contributions
from the ID PMTs themselves.
The actual size of the current water tank is therefore large
enough to house future extensions of the XMASS experiment.
The water is constantly circulated at ${\rm 5 \, ton/h}$ and purified 
through a system consisting of filters, an ion-exchanger, 
a UV sterilizer and a membrane degasifier 
to remove impurities and radon gas.
Radon reduced air processed by charcoal is 
supplied to the space between the water surface and 
top of the water tank. 

\section{ Structure of ID  }

The detector employs single-phase liquid xenon and 
uses only scintillation light 
emitted by dark matter or other interactions.
No electric field is necessary as no ionisation signal is extracted 
from the liquid.
This makes it possible to realize a large scale detector 
and maximize the detector uniformity and volume-to-surface ratio.
Pulse shape discrimination for single-phase liquid xenon is
also being studied \cite{PSD}.

\begin{figure}
\begin{center}
\includegraphics[width=12.0cm]{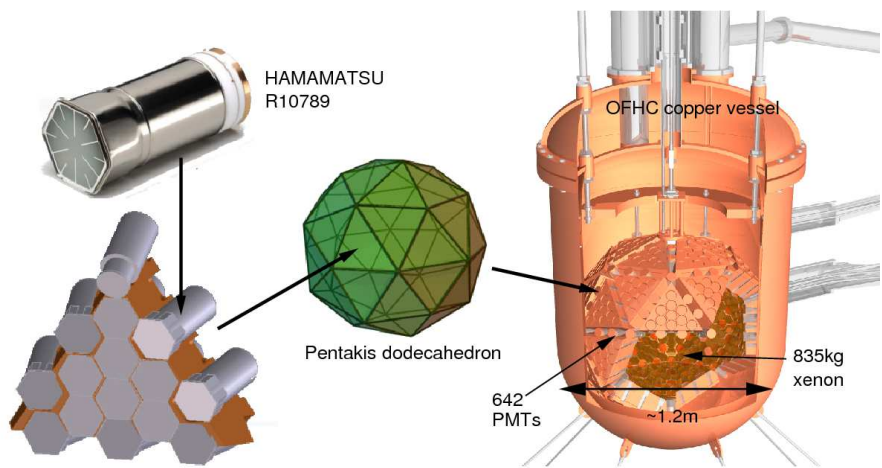}
\caption{Structure of the ID.}
\label{fig:fig_detector}
\end{center}
\end{figure}

Figure \ref{fig:fig_detector} shows the structure of the ID.
Six hundred and thirty hexagonal PMTs (HAMAMATSU R10789-11)
and 12 round PMTs (HAMAMATSU R10789-11MOD)
are mounted in an oxygen free high conductivity (OFHC) 
copper holder with an approximately spherical shape 
called a pentakis dodecahedron.
Because the photo-cathode area is maximized using hexagonal PMTs,
a photo coverage of more than ${\rm 62\%}$ is achieved.
The quantum efficiency at the scintillation wavelength 
of liquid xenon ($\sim$${\rm 175 \, nm}$)
is more than ${\rm 28\%}$.
The shape of the pentakis dodecahedron consists of 60 isosceles triangles.
One triangular module holds approximately 10 PMTs.
The entire structure is immersed in liquid xenon.
The vessel consists of outer and inner OFHC copper vessels
${\rm 1,280 \, mm}$ and ${\rm 1,120 \,mm}$ in diameter, respectively.
The inner vessel holds liquid xenon and the PMT holder, 
and the outer vessel is used for vacuum insulation.
To reduce the amount of liquid xenon, 
an OFHC copper filler is installed in the gap 
between the PMT holder and tne inner vessel.

To avoid cosmogenic activation,
the copper material used for the PMT holder and vessels 
was brought into the Kamioka mine
within one month after it was electrorefined 
by Mitsubishi Materials Co., Ltd. 
From there it was sent to the company
where the PMT holder was made,
and then the holder was sent to the Kamioka mine.
The total exposure time of copper on the surface 
was less than 130 days.
This way, the production of ${\rm ^{60}Co}$, which is one of 
the cosmogenic isotopes in copper \cite{cosmogenic}, 
was kept below our required level (${\rm 0.25 \, mBq/kg}$).
The estimated BG from ${\rm ^{60}Co}$ in copper is negligible compared
with our main BG because of the PMTs.
To clean the surface after processing the PMT holder was chemically etched,
while the inner surface of the inner vessel holding the xenon was
electropolished, both to reduce radon and hydrogen emanation and 
$^{210}$Pb contamination.

PMT mounting and detector assembly were performed inside a clean booth
constructed in the water tank from December 2009 to September 2010.
During the work, the radon concentration in the water tank was 
maintained at approximately ${\rm 200 \, mBq/m^{3}}$,
which is 1/100 of that in the atmosphere, 
and the dust level was less than 
${\rm 1000 \, particles/ft^{3}}$ (particle size ${\rm > 0.5 \, \mu m}$).

\section{ Distillation system }

Xenon does not have long-lived radioactive isotopes, which is one of 
its most important advantages for a rare event search 
such as a dark matter search.
However, commercial xenon contains a small amount (0.1--1${\rm \, ppm}$) 
of krypton, which has a radioactive isotope, 
$\rm ^{85}Kr$ (half-life of 10.76 years).
The xenon acquired for the XMASS detector had 
a Kr/Xe content of ${\rm 340 \, ppb}$, and a ${\rm ^{85}Kr/Kr}$
ratio of ${\rm (0.6\pm0.2)\times 10^{-11}}$.
This ${\rm ^{85}Kr/Kr}$ ratio is an actual measured value.
After the distillation, xenon with high Kr concentration was 
obtained as off-gas, which enabled us to measure ${\rm ^{85}Kr}$
using a high purity germanium (HPGe) detector.
Our requirement for Kr/Xe is less than ${\rm 2 \, ppt}$.
Together with a company we developed a prototype distillation system that
achieved a Kr/Xe value of $\rm 3.3 \pm 1.1 \, ppt$; 
the details are reported in \cite{distillation}.
In order to process ${\rm 1.2 \, ton}$ of xenon for the XMASS detector, 
a new, high throughput system was built.
The new distillation system has a tower length of 
${\rm 2,915 \, mm}$ and column diameter of ${\rm 40 \, mm}$.
It is designed to reduce the amount of Kr by five orders 
of magnitude with one pass.
The system can collect purified xenon with $99 \%$ efficiency.
The system had a process flow rate of ${\rm 4.7 \, kg/h}$, and 
${\rm 1.2 \, ton}$ of xenon was processed in 10 days 
before it was introduced into the detector. 
During the commissioning run,
a sample was taken and its krypton concentration was measured 
by atmospheric pressure ionization mass spectroscopy (API-MS). 
No excess over the BG level of the API-MS was found, and
an upper limit of 2.7 ppt was obtained for its krypton contamination.

\section{ Xenon circulation and purification system}

\begin{figure}
\begin{center}
\includegraphics[width=12.0cm]{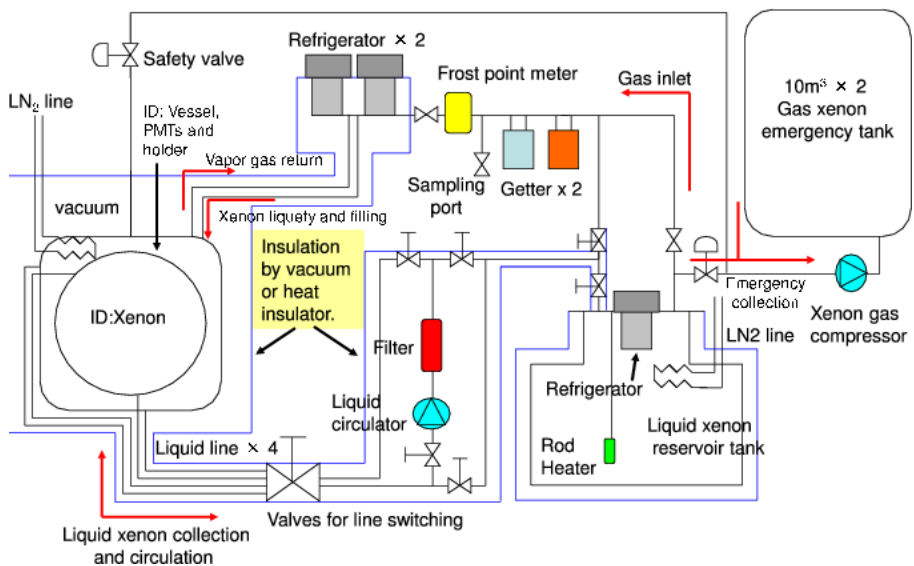}
\caption{ Flow map of xenon circulation and purification system for the XMASS detector.}
\label{fig:fig_plumbing}
\end{center}
\end{figure}

Figure \ref{fig:fig_plumbing} shows the flow map of the xenon circulation and
purification system for the XMASS detector.
Xenon was kept in the liquid xenon reservoir tank before the experiment.
A ${\rm 200 \, W}$ cooling power pulse tube refrigerator (PC105U, IWATANI) 
and vacuum insulation held the liquid xenon. 
The heat leakage was less than ${\rm 50 \, W}$. 
A rod heater in the liquid can evaporate the xenon.

When we filled the ID with liquid xenon, xenon was evaporated in the
reservoir tank using the rod heater at approximately ${\rm 300 \, W}$.
The evaporated gas was purified by two getters (PS4-MT15, SAES),
which are connected in series, at a flow rate of  ${\rm 30 \, L/min}$.
The first stage getter was operated at high temperature, as designed,
but the second one was operated at room temperature to obtain 
better performance for removing hydrogen impurities.
The purified gas was monitored by a frost point meter 
(Pura, Michell Instruments),
and the measured value was less than ${\rm -120 ^{\circ}C}$ 
at ${\rm 0.06 \, MPa}$, which
corresponded to less than ${\rm 0.09 \, ppb}$ water in the gas.
Before the ID was filled with liquid xenon, 
it was cooled by liquid 
nitrogen to ${\rm -100 ^{\circ}C}$, and the xenon gas was liquefied by two
pulse refrigerators (PC105U, IWATANI) each having
a cooling power of ${\rm 200 \, W}$.
It takes five days to fill the ID with liquid xenon in this way.

After the first filling, we collected the liquid xenon from the ID
to the reservoir tank at a liquid flow rate of ${\rm 2 \, L/min}$. 
We completed the xenon collection in approximately ${\rm 3 \, h}$.
This empty-refill procedure with purification was repeated two times,
and an increase in light yield of approximately $16\%$ 
compared with that obtained after the first filling 
was observed with a $\rm ^{57}Co$ source.
The radon concentration in the liquid xenon was evaluated from data recorded
in the XMASS detector.
The obtained values were ${\rm 8.2\pm0.5 \, mBq}$ for ${\rm ^{222}Rn}$
and $<$${\rm 0.28 \, mBq}$ for ${\rm ^{220}Rn}$.
Although the BG level caused by radon was estimated to be 
sufficiently low (${\rm \sim 10^{-4} \, keV^{-1}kg^{-1}day^{-1}}$ at ${\rm 5 \, keV}$) by a simulation study,
further study for reduction of radon continues \cite{motoki}.

One of the two pulse refrigerators is used to maintain the liquid xenon
temperature in the ID during normal operation.
Less than ${\rm 35 \, W}$ of cooling power are needed to compensate for heat leakage.
Should the xenon gas pressure above the liquid surface in the ID
change for some reason safety system will automatically be activated.
The first system uses liquid nitrogen to reduce the temperature in the ID
in case e.g.\ the electrically powered refrigerator fails.
The second system collects the xenon by compressing the evaporating gas
into two ${\rm 10 \, m^{3}}$ emergency gas tanks.
All gas lines are protected by a appropriately chosen rupture disk.

We can also circulate and purify the liquid xenon in the ID using a liquid xenon
circulator and filter. The circulation speed is ${\rm 2 \, L/min}$ as liquid.
Liquid lines are connected from the inside/outside and 
top/bottom sides of the PMT holder 
to the circulator. The circulation lines can be selected by valve operation.

\section{ Low-BG PMT }
Because gamma rays and high-energy neutrons from the outside
are shielded against by the OD,
the major origin of the BG was expected to be gamma rays 
from radioactive impurities inside the ID PMTs.
To minimize the BG, we developed low-radioactivity PMTs 
(R10789-11 and R10789-11MOD) with Hamamatsu Photonics K.K. 
giving special attention to radioactive contaminants in their components.
Table \ref{tab:pmt} shows the radioactive
contaminants in a PMT, including its voltage divider base as evaluated
with HPGe detectors.

During the commissioning run further measurements for
Kovar alloy (body of PMT) and aluminium components (sealing
between the PMT body and its window) by HPGe were made to estimate
their contribution of low-energy gamma rays and related 
beta rays in addition to high-energy gamma rays
from radioactive isotopes.
Because Kovar and aluminium components are in contact with the liquid xenon 
of the sensitive region,
beta rays from a radioactive isotope possibly contribute to the BG.
Although a quartz window is also in contact with the liquid xenon,
BG events occurring in the front of the window from its contaminants
can be easily rejected using information of hit pattern of PMTs.
Because we cannot measure the Kovar and aluminium components currently with the XMASS detector,
we measured samples of the same materials produced by the same company.
Table \ref{tab:kovar} and \ref{tab:alumi} shows radioactive
contaminants in Kovar and aluminium samples, respectively.
Significant amounts of ${\rm ^{234}Th}$ and ${\rm ^{210}Pb}$
were found in aluminium samples.

Three samples were used for the measurements of aluminium components.
Sample 1 consists of leftovers from the actual PMT production; 
$\sim$${\rm 15\%}$ of the PMTs in the XMASS detector were fabricated
with materials from this particular batch.
To evaluate variability of contaminants between batches,
we also measured a sample 2 with material from the prototype production for our PMTs,
and finally a sample 3 which is a bulk sample from the same supplier as the 
previous samples.
The results show the existence
of the contaminants ${\rm ^{234}Th}$, ${\rm ^{210}Pb}$ and ${\rm ^{232}Th}$ 
inside the aluminium components used for the PMTs although the
contaminants of ${\rm ^{214}Bi}$ in the same Uranium chain is much smaller \cite{LRT}. 
Concentrations of Uranium and Thorium in sample 1 were confirmed 
by Hamamatsu Photonics K.K. using inductively coupled plasma-mass spectrometry (ICP-MS) 
as shown in Table \ref{tab:icp-ms}.
The contributions of these materials to the observed
BG will be described elsewhere.

\begin{table}[hbtp]
\begin{center}
\caption{Radioactive contaminants in the hexagonal PMT (R10789-11) 
and its base. All values are in units of ${\rm mBq/PMT}$.
Decay gamma rays were measured with HPGe detectors.
The observed gamma ray spectrum was analysed using efficiencies 
evaluated by Monte Carlo simulation.
To evaluate the Uranium and Thorium chain, 
gamma rays from ${\rm ^{214}Pb}$ and ${\rm ^{214}Bi}$,  and
 ${\rm ^{228}Ac}$, ${\rm ^{212}Pb}$, ${\rm ^{212}Bi}$ and ${\rm ^{208}Tl}$, were used.}
{\begin{tabular}
{lc}
\hline
Isotopes &  Radioactivity [mBq/PMT]\\
\hline
${\rm ^{214}Pb}$, ${\rm ^{214}Bi}$ (U-chain)   &  ${0.70  \pm 0.28}$ \\
${\rm ^{228}Ac}$, ${\rm ^{212}Pb}$, ${\rm ^{212}Bi}$, ${\rm ^{208}Tl}$ (Th-chain) &  ${1.5   \pm 0.31}$ \\
${\rm ^{40}K}$      &  ${ < 5.1}$ \\
${\rm ^{60}Co}$     &  ${2.9   \pm 0.16}$ \\
\hline
\end{tabular}}
\label{tab:pmt}
\end{center}
\end{table}

\begin{table}[hbtp]
\begin{center}
\caption{Radioactive contaminants in the Kovar sample. All values are in units of ${\rm mBq/PMT}$.
Isotopes for the Thorium chain are the same as those in Table 1.
Note that only a small fraction of the Kovar body is protruding into the ID volume through.}
{\begin{tabular}
{lc}
\hline
Isotopes  &  Kovar sample [mBq/PMT]\\
\hline
${\rm ^{234}Th}$  & ${ < 2.2}$  \\
${\rm ^{214}Bi}$  & ${ < 0.4}$  \\
${\rm ^{210}Pb}$  & ${ < 6.4}$ \\
Th-chain & ${ < 0.32}$ \\
\hline
\end{tabular}}
\label{tab:kovar}
\end{center}
\end{table}

\begin{table}[hbtp]
\begin{center}
\caption{Radioactive contaminants of various aluminium samples.
Leftover material from the actual PMT production (sample 1), 
material from the prototype production (sample 2) 
and a bulk sample (sample 3). 
Sample 2a and 2b shows measurement with low and high statistics 
before and after detector construction, respectively.
All values are in units of ${\rm mBq/PMT}$.
Decay gamma rays were evaluated by measurements 
with HPGe detectors as described in Table 1.Isotopes for the Thorium chain are the same as those in Table 1.}
{\begin{tabular}
{lcccc}
\hline
Isotopes  &  Sample 1 & Sample 2a & Sample 2b & Sample 3\\
\hline
${\rm ^{234}Th}$  & ${1.1  \pm 0.12}$         &  ${1.0 \pm 1.1}$ & ${0.33 \pm 0.36}$ & ${ 1.8 \pm 0.23 }$ \\
${\rm ^{214}Bi}$  & ${ < 1.7 \times 10^{-2} }$ &  $< 1.0 \times 10^{-1}$ & $ < 9.1 \times 10^{-2}$ & $ < 3.6 \times 10^{-2}$ \\
${\rm ^{210}Pb}$  & ${ 2.6 \pm 2.0 }$         &  NA              & ${8.7 \pm 6.3}$ & ${ 7.7 \pm 3.3  }$  \\
Th-chain  & ${ (3.5 \pm 1.0)\times 10^{-2} }$ &  $< 1.3 \times 10^{-1}$   & $< 5.4 \times 10^{-2}$ & ${ (1.8 \pm 0.2)\times 10^{-1}  }$  \\
\hline
\end{tabular}}
\label{tab:alumi}
\end{center}
\end{table}

\begin{table}[hbtp]
\begin{center}
\caption{Concentration of Uranium and Thorium in sample 1 measured by ICP-MS.
Concentrations of ${\rm 0.31 \, \mu g/g}$ for Uranium and 
${\rm 0.018 \, \mu g/g}$ for Thorium correspond to ${\rm 1.1 \, mBq/PMT}$
and ${\rm 2.2 \times 10^{-2} \, mBq/PMT }$, respectively.
}
{\begin{tabular}
{lc}
\hline
Elements  &  Sample 1 [${\rm \mu g/g}$ ]\\
\hline
${\rm U}$   & ${\rm 0.31 \pm 0.03}$   \\
${\rm Th}$  & ${\rm 0.018 \pm 0.002}$  \\
\hline
\end{tabular}}
\label{tab:icp-ms}
\end{center}
\end{table}

\section{ Material selection }
All the components and materials used for the detector were carefully vetted.
More than 250 potential candidates for components were measured
with the HPGe detectors.
The criterion for material selection was that based on our simulation
no material should contribute more than a small fraction of the BG
expected from the PMTs.

The radon emanation from components inside the ID was also measured by
a highly sensitive radon detector \cite{radon_detector} before
construction.
After detector construction, the radon concentration in the detector's liquid xenon
was evaluated by analysing the data from the XMASS detector.

Impurity elution from the components was also checked beforehand 
in terms of the degradation of the light yield.
For this we used a small
test chamber consisting of two PMTs (R10789-11) and $\sim$${\rm 3 \, kg}$
of liquid xenon.

\section{ Data acquisition system }

\begin{figure}
\begin{center}
\includegraphics[width=10.0cm]{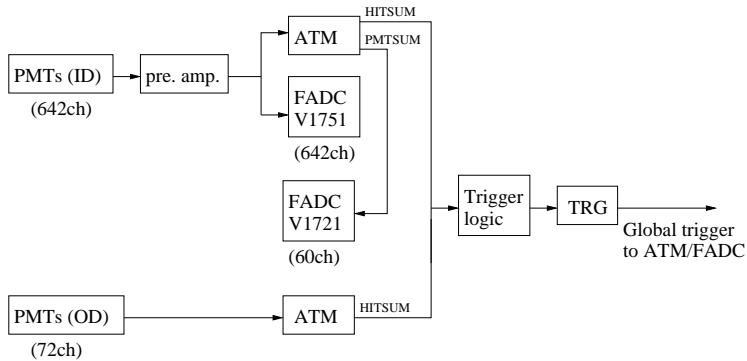}
\caption{Schematic view of the data acquisition system.}
\label{fig:fig_daq}
\end{center}
\end{figure}

Figure \ref{fig:fig_daq} shows a schematic diagram of the data acquisition system.
Six hundred and forty two PMT signals from the ID are sent to the electronics hut located 
on top of the water tank.
The signals go through $\sim$${\rm 11 \, m}$ of coaxial cables
and feed-throughs (resin type, RHS) that separate the xenon gas volume from the outside air.
In the electronics hut, the signals are processed and digitized.
The signals are amplified by a factor of 11 using preamplifier cards and are then fed into 
analog-timing-modules (ATMs) \cite{ATM} as well as CAEN V1751 flash analogue-to-digital converter (FADC) inputs.
The ATMs were originally made for and used in Super-Kamiokande I-III and 
are reused in the XMASS detector.
They function as ADCs and time-to-digital converters (TDCs), and
record the integrated charge and arrival time of each
PMT signal.
The dynamic ranges are approximately ${\rm 450 \, pC}$ (corresponding to approximately 120 
photo-electrons (PEs))
with a resolution of ${\rm 0.2 \, pC}$ (corresponding to ${\rm 0.05 \, PEs}$) for the ADC and 
approximately ${\rm 1300 \, ns}$ with a resolution of ${\rm 0.4 \, ns}$ for the TDC.
Ten or eleven PMTs, most of which are mounted on the same triangular holder, are connected
to one ATM board.
CAEN V1751 FADCs (${\rm 1 \, GHz}$ sampling rate) were installed 
in December 2011 for better understanding of the BG and more sophisticated
searches for dark matter signals using timing information.
The dynamic range is ${\rm 1 \, V}$ with ${\rm 10 \, bit}$ (${\rm 1 \, mV}$) 
resolution (${\rm 1 \, mV}$ corresponding to ${\rm 0.05 \, PE}$).
The bandwidth is ${\rm 500 \, MHz}$.
Seventy two PMT signals from the OD are also processed and digitized in the electronics hut
and recorded by ATMs. 
Pedestal data are taken every 5 min to minimize the effect
of possible drifts due to changes in temperature.
This accounts for most of the dead time, which is 1--2$\%$.

When the ATM input signal is less than ${\rm -5 \, mV}$, which corresponds to 
approximately ${\rm 0.2 \, PEs}$ in the ID  and ${\rm 0.4 \, PEs}$ in the OD, 
a rectangular signal ${\rm 200 \, ns}$ wide and ${\rm 15 \, mV}$ high
is generated.
An output signal on the ATM front panel called HITSUM is generated
by summing up all the rectangular analogue signals generated on channels belonging to that ATM module.
All the HITSUM signals from the ID ATMs are summed up again by a summing amplifier
to generate the global ID HITSUM signal. 
When the global ID HITSUM signal is higher than the ID trigger 
threshold, an ID trigger is issued.
The ID trigger threshold was at the nine hits until September 28, 2011, 
and thereafter, at the three hits.
A separate trigger for the OD is generated in an analog manner from OD signals on the OD ATM boards.
The OD trigger threshold is at the seven hits.
Typical trigger rates for the ID and OD are $\sim$${\rm 4 \, Hz}$
and $\sim$${\rm 7 \, Hz}$, respectively.
The ID and OD trigger signals, as well as external clock trigger signals,
are fed into a custom-made VME trigger module (TRG), which was also used for 
Super-Kamiokande I-III.
For any type of trigger input, the TRG generates a global trigger signal
and a 16-bit event number and sends them to both the ATM and FADC to 
initiate data collection for each event.
The TRG module records the trigger time with a resolution of ${\rm 20 \, ns}$,
together with the trigger type and event number.
After an ID trigger is issued, a ${\rm 6 \, \mu s}$ trigger veto is applied because
many after pulses due to high energy events would issue unnecessary triggers.

The ATM outputs the PMTSUM signal, which is an analogue sum of the 12 channels 
in one board, after the signal is attenuated by a factor of eight compared to the
input signal. 
Because we have 60 ATM boards, 60 PMTSUM signals are fed into some older CAEN V1721 FADCs, which have
a ${\rm 500 \, MHz}$ sampling rate. 
We started recording these waveforms in February 2011.
The dynamic range of these FADCs is ${\rm 1 \, V}$ with ${\rm 8 \, bit}$ (${\rm 4 \, mV}$) 
resolution, corresponding to a ${\rm 1.25 \, PE}$
pulse height. The acquisition window is ${\rm 4 \, \mu s}$ (approximately ${\rm 1 \, \mu s}$ 
before the trigger and ${\rm 3 \, \mu s}$ after it).
The FADC data is primarily used for understanding the BG.

\section{ Status monitor }
The status of the detector and its environment is monitored to maintain stable 
data acquisition and safe operation. 
Many sensors for temperature, pressure, liquid xenon level, 
cryostat vacuum and PMT voltages and currents, etc.\ are monitored.
Table \ref{tab:slowmon} summarizes the number of sensor channels.
Readings from the sensors related to the liquid xenon status are examined every second
by a web-based alert system.
Two shift workers who are always present at the laboratory
also cross-check the status using on-site display instruments every hour.
The pressure and temperature of the detector in normal operation were
${\rm 0.165 \, MPa \, (abs.)}$ and ${\rm -99 ^{\circ}C}$, respectively.
Except for some special data acquisition periods, the pressure change in the ID
was within ${\rm 0.006 \, MPa}$, and the temperature change in the ID was 
within ${\rm 1.2 ^{\circ}C}$.

\begin{table}
\begin{center}
\caption{Number of sensor channels in the status monitor system.}
{\begin{tabular}
{lc}
\hline
Items  &  Number of channels \\
\hline
Temperature                &  117  \\
Pressure                   &   18  \\
Liquid xenon level         &    3  \\
Vacuum                     &    4  \\
Refrigerator heater duty   &    5  \\
Water purification         &   22  \\
Radon level                &    8  \\
ID HV                      &  648  \\
OD HV                      &   72  \\
Others                     &   57  \\
\hline
Total                      &  954 \\
\hline
\end{tabular}}
\label{tab:slowmon}
\end{center}
\end{table}

The PMT gain stability is monitored with signals generated by blue LEDs in the ID. 
A total of eight LEDs with Teflon diffusers are
embedded in holes on the surface of the ID PMT holder.
One PE signals were taken with a low
hit occupancy every week, and the ADC mean was used to check stability. 
The PMT gain is stable within $\pm \, 5\%$.

\section{ Commissioning run }
During the commissioning period, various types of data were taken in addition to normal run data 
for the dark matter search. 
Calibration runs using radioactive sources for better modelling of the detector in a 
simulation are summarized in the next section.
The runs in Table \ref{tab:runsum} were used for optimisation of the detector operation
for the dark matter search
and further understanding of the BG.
High- and low-pressure runs using O$_2$-doped liquid xenon (O$_2$ injection run) and 
heat injection to produce higher fluctuations in the density of liquid xenon (boiling run) were
made to explore the liquid xenon's optical properties.
A gas run was made to understand the BG from the ID's inner surface. 
For this run, the ID was filled with xenon gas instead of liquid.
These runs were useful for understanding various conditions of the detector
as well as the BG.
The details on them will be described elsewhere.
Science results from the commissioning run are presented in Ref.\ \cite{LowMassWIMP, Axion}.
\begin{table}
\begin{center}
\caption{Special commissioning run summary.}
{\begin{tabular}
{lc}
\hline
Run condition       &  Operation term\\
\hline
High-pressure run (${\rm 0.23 \, MPa \, (abs.)}$) & From March 25 2011 to April 10 2011 \\
Low-pressure run (${\rm 0.13 \, MPa \, (abs.)}$)  & From April 13 2011 to May 4 2011 \\
O$_2$ injection run          & From September 28th, 2011 to January 11th, 2012 \\
Boiling run                  & From January 4 2012 to January 10 2012 \\
Gas run                      & From January 13 2012 to January 24 2012 \\
\hline
\end{tabular}}
\label{tab:runsum}
\end{center}
\end{table}

\section{ Calibration }
\begin{figure}
\begin{center}
\includegraphics[width=11.0cm]{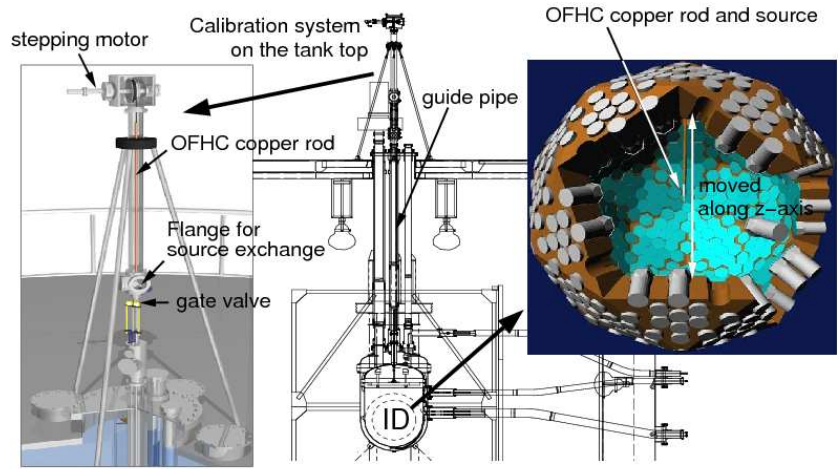}
\caption{Calibration system on top of the tank. Source placed on the edge of the copper rod is inserted into the ID and can be moved along the z axis.}
\label{fig:fig_calib}
\end{center}
\end{figure}

A calibration system is connected to the ID. This system enables us to
drive a radioactive source inside the ID PMT holder along 
the vertical axis (z axis) with a precision of better than 
${\rm 1 \, mm}$.
Figure \ref{fig:fig_calib} shows the calibration system, which consists of 
a radioactive source, 
an OFHC copper rod, a thin stainless steel wire and a stepping motor.
The radioactive source can be mounted at the tip of the OFHC copper rod.
The rod is hung by the thin stainless steel wire, lowered through
a guide pipe and inserted into the ID using a motion feed-through 
and stepping motor on top of the water tank.

The radioactive source can be exchanged without interrupting 
detector operation.
The source and their corresponding energies are 
summarized in ${\rm Table \, \ref{tab:source}}$.
The ${\rm ^{241}Am}$ and ${\rm ^{57}Co}$ have a diameter
of ${\rm 0.21 \, mm}$, which is much smaller than 
the absorption lengths for ${\rm 59.5}$ and ${\rm 122 \, keV}$ gamma rays 
in liquid xenon ($\sim$$\rm 0.44 \, mm$ and $\sim$$\rm 2.5 \, mm$, respectively).
This reduces the shadow cast by the source assembly in the scintillation light pattern
of these sources.
All enclosure assemblies used in the XMASS detector so far are cylindrical.

\begin{table}
\begin{center}
\caption{Calibration sources and energies. 
The 8 keV (*1) in the $\rm ^{109}Cd$ and 
59.3 keV (*2) in the $\rm ^{57}Co$ source are $\rm K_{\alpha}$ X-rays 
from the copper and tungsten, respectively,
used for source housing. }
{\begin{tabular}
{lrl}
\hline
  Isotopes     &  Energy [keV]  & Shape\\
\hline
${\rm ^{55}Fe}$ & 5.9  & cylinder \\
${\rm ^{109}Cd}$ & 8(*1), 22, 58, 88 & cylinder\\
${\rm ^{241}Am}$ & 17.8, 59.5 & thin cylinder\\
${\rm ^{57}Co}$ & 59.3(*2), 122 & thin cylinder\\
${\rm ^{137}Cs}$ & 662 & cylinder\\
\hline
\end{tabular}}
\label{tab:source}
\end{center}
\end{table}

There is also an external calibration system.
A U-shaped soft hose was attached to the outside of the vessel.
A radioactive source can be moved around the detector through the hose.
This makes it possible to study the detector response to external radiation.

\section{ Detector simulation }
An XMASS detector Monte Carlo simulation (MC) package 
based on Geant4 \cite{geant4}
has been developed.
The particle tracks, scintillation process, 
propagation of scintillation photons, 
PMT response and readout electronics are simulated by this MC package.

\begin{figure}
\begin{center}
\includegraphics[width=10.0cm]{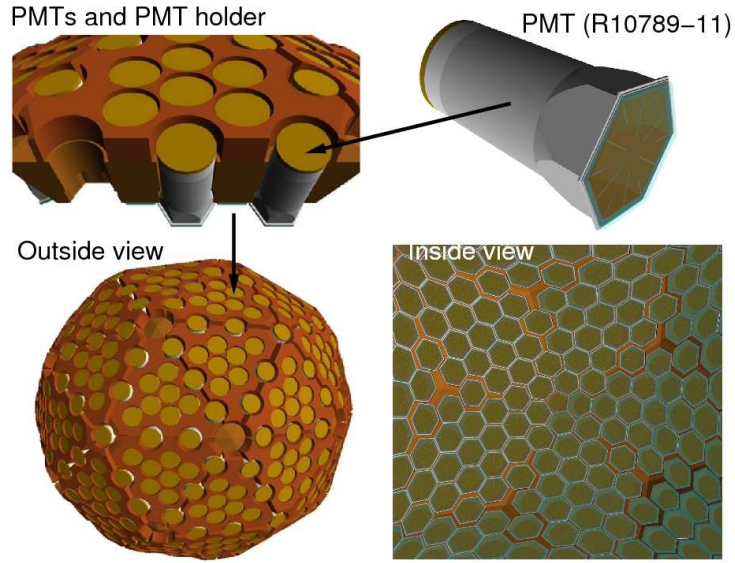}
\caption{ID geometry, excluding vessels, in MC. Upper right: geometry of one hexagonal PMT. 
Upper left: part of PMT holder and some PMTs. Bottom: entire PMT holder and PMTs: outside and inside view.}
\label{fig:fig_detector_mc}
\end{center}
\end{figure}

\begin{figure}
\begin{center}
\includegraphics[width=8.5cm]{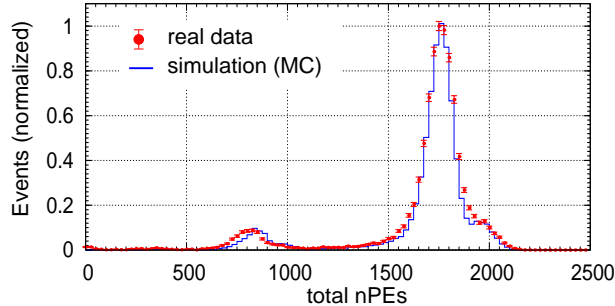}
\caption{Observed $n$PE spectrum using the $\rm ^{57}Co$ source at ${\rm z = 0 \, cm}$ (red dots). Simulated spectrum is shown as blue histogram.}
\label{fig:fig_totalpe_co57}
\end{center}
\end{figure}

Figure \ref{fig:fig_detector_mc} shows the geometry of the ID 
in the XMASS MC.
The copper vessel is also defined, but it is omitted in the figure.
For tracing each scintillation photon, 
the precise geometries and optical properties of 
all the components in contact with liquid xenon are defined.
Several optical parameters need to be determined experimentally.
They are the absorption and scattering coefficients of liquid xenon,
reflectance at the inner surface of the PMT copper holder 
and the aluminium strip on the PMT window,
refractive indices of liquid xenon and the quartz PMT window,
and the reflection and absorption probabilities at a PMT photo-cathode.
These parameters were tuned so that the observed numbers of PEs ($n$PEs)
in each PMT 
in the simulated samples reproduce those 
in the data for various source positions. 
Figure \ref{fig:fig_totalpe_co57} shows the $n$PE spectrum 
observed using the $\rm ^{57}Co$ source at ${\rm z = 0 \, cm}$ and
the MC result.
The $n$PE distribution was reproduced well by the MC, 
and a high light yield, $14.7 \pm 1.2$ PE/keV was
obtained.    

The MC tracks the incident particles and any energy deposited through various interactions.
From the deposited energies in each vertex, 
scintillation photons are generated by
taking into account the dependence on energy and nature of the depositing particle,
implementing a realistic, non-linear scintillation efficiency \cite{scinti_eff}.
This effect results in non-linearity of the scintillation efficiency.
The energy distribution of the scintillation photons is based on 
the measured value \cite{scinti_ene}
which is a Gaussian distribution with a mean value of ${\rm 7.078 \, eV}$ 
and standard variation of ${\rm 0.154 \, eV}$.  
The propagation of scintillation photons is simulated and photons 
absorbed at the photo-cathode of the PMT are treated 
as detected PEs, taking into account the quantum efficiency of each PMT.
The response of each PMT for single photon detection is also simulated
on the basis of a single photon distribution measured 
using low-intensity LEDs.

\begin{figure}
\begin{center}
\includegraphics[width=8.0cm]{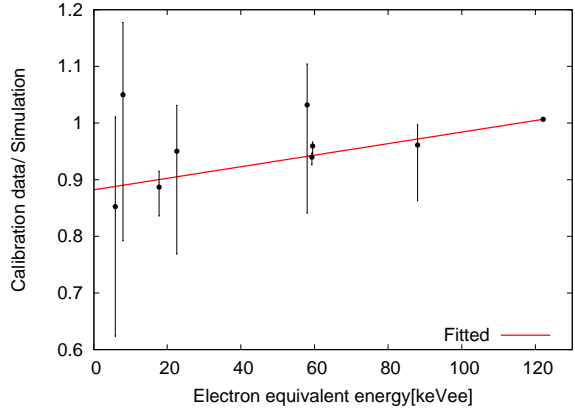}
\caption{Comparison of energy scales in calibration data and MC.}
\label{fig:fig_energyscale}
\end{center}
\end{figure}

Although our MC considers the non-linearity of the scintillation efficiency, 
some deviations exist in the total $n \rm PE$ from data 
obtained for gamma rays with energies less than 122 keV.
Figure \ref{fig:fig_energyscale} shows the 
ratios of the observed $n \rm PE$ and predicted $n \rm PE$ by the MC.
This deviation is treated as the systematic error 
at the energy scale of our detector.

\section{ Vertex reconstruction }
The vertex positions and energies of events were 
reconstructed using $n \rm PE$ information 
from the PMTs.
For various grid positions inside the ID, expected $n \rm PE$ distributions
in each PMT were calculated in the MC.
We use positions on a Cartesian grid, on radial lines 
from the centre of the detector,
and on the inner surface of the detector including gaps between PMTs.
These distributions are normalized so that they can be used
as the probability density functions (PDFs) for each grid position.
The probability, $p_{i}(n)$ that the i-th PMT detects 
$n \rm PE$ is calculated
using the PDF.
The likelihood that the vertex is in the assumed position 
$\mathbf x$ is the product
of all $p_{i}(n_{i})$;
\begin{equation}
L({\mathbf x}) = \prod_{i=1}^{642} p_{i}(n_{i}),
\end{equation}
where $n_{i}$ represents $n \rm PE$ for the i-th PMT.
The most likely position is obtained by maximizing $L$.

\begin{figure}
\begin{center}
\includegraphics[width=9.0cm]{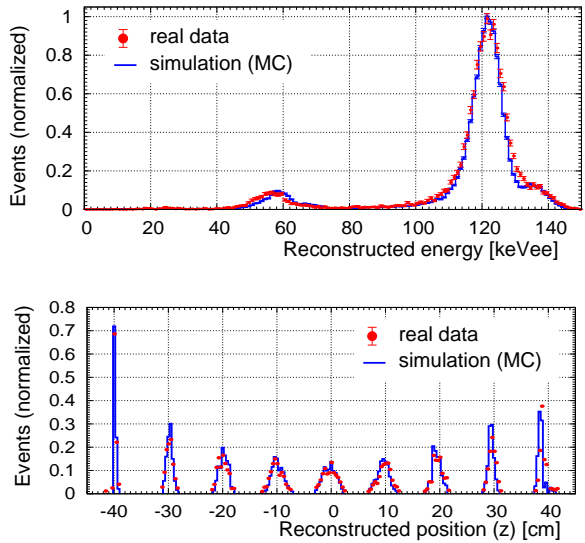}
\caption{Energy spectra reconstructed using the
$\rm ^{57}Co$ source at ${\rm z = 0 \, cm}$ (upper) and 
vertex distributions reconstructed
using the same source at ${\rm z = -40,\, -30,\, ..., 40 \, cm}$ (lower).}
\label{fig:fig_co57}
\end{center}
\end{figure}

The performance of the vertex and energy reconstruction 
was evaluated using several types of radioactive sources.
The upper panel of Fig.\ \ref{fig:fig_co57} shows 
the energy spectrum reconstructed
using same data set in Fig.\ \ref{fig:fig_totalpe_co57}.
The energy resolution for ${\rm 122 \, keV}$ gamma rays is $\rm 4 \%$ (r.m.s).
The lower panel of Fig. \ref{fig:fig_co57} shows the reconstructed vertices
for various $\rm ^{57}Co$ source positions.
The observed position resolution (rms.) 
is ${\rm 1.4 \, cm}$ at ${\rm z = 0 \, cm}$ and
${\rm 1.0 \, cm}$ at ${\rm z = \pm 20 \, cm}$ for ${\rm 122 \, keV}$ gamma rays.
The distributions of the reconstructed energy 
and vertices for ${\rm 122 \, keV}$ gamma rays
are reproduced well by the MC.

\section{Conclusion}
The construction of the XMASS detector was completed in September 2010 and
commissioning runs were conducted from October 2010 to June 2012.

The XMASS detector is the world's largest (ton scale) single-phase liquid 
xenon detector for dark matter searches.
The key idea for BG reduction is self shielding using vertex reconstruction.
The position and energy resolution along the z axis inside the detector 
were measured 
with radioactive sources and are well reproduced by MC.
The observed position and energy resolution 
for ${\rm 122 \, keV}$ gamma rays are ${\rm 1.0 \, cm}$ 
at ${\rm z = \pm 20 \, cm}$ and $4 \%$ (r.m.s) at ${\rm z = 0 \, cm}$, respectively.

A high light yield, $14.7 \pm 1.2$ PE/keV was obtained
owing to the large photocoverage ($>$$62\%$) and small 
amount of impurities in the liquid xenon.
This was achieved by careful control of dust and radon 
during construction and purification by liquid collection 
and filling with purified gas.
The concentrations of radon (${\rm 8.2\pm0.5 \, mBq/835kg}$ for 
${\rm ^{222}Rn}$ and $<$${\rm 0.28 \, mBq/835kg}$ 
for ${\rm ^{220}Rn}$) and krypton ($<$$\rm 2.7 \, ppt$) in liquid xenon 
are also the lowest among liquid xenon detectors for dark matter searches.

\section*{Acknowledgements}
We gratefully acknowledge the co-operation of Kamioka Mining
and Smelting Company. 
This work was supported by the Japanese Ministry of Education,
Culture, Sports, Science and Technology, Grant-in-Aid
for Scientific Research, and partly supported
by the National Research Foundation of Korea Grant funded
by the Korean Government (NRF-2011-220-C00006).





\bibliographystyle{model1-num-names}
\bibliography{<your-bib-database>}



\end{document}

%% file: tech-auth.tex
\author[ICRR,IPMU]{K.~Abe}
\author[ICRR]{K.~Hieda}
\author[ICRR,IPMU]{K.~Hiraide}
\author[ICRR]{S.~Hirano}
\author[ICRR,IPMU]{Y.~Kishimoto}
\author[ICRR,IPMU]{K.~Kobayashi}
\author[ICRR,IPMU]{S.~Moriyama}
\author[ICRR]{K.~Nakagawa}
\author[ICRR,IPMU]{M.~Nakahata}
\author[ICRR]{H.~Nishiie}
\author[ICRR,IPMU]{H.~Ogawa}
\author[ICRR]{N.~Oka}
\author[ICRR,IPMU]{H.~Sekiya}
\author[ICRR]{A.~Shinozaki}
\author[ICRR,IPMU]{Y.~Suzuki}
\author[ICRR,IPMU]{A.~Takeda}
\author[ICRR]{O.~Takachio}
\author[ICRR]{K.~Ueshima\fnref{tohoku}}
\author[ICRR]{D.~Umemoto}
\author[ICRR,IPMU]{M.~Yamashita}
\author[ICRR]{B.~S.~Yang}

\author[GIFU]{S.~Tasaka}

\author[IPMU]{J.~Liu}
\author[IPMU]{K.~Martens}

\author[KOBE]{K.~Hosokawa}
\author[KOBE]{K.~Miuchi}
\author[KOBE]{A.~Murata}
\author[KOBE]{Y.~Onishi}
\author[KOBE]{Y.~Otsuka}
\author[KOBE,IPMU]{Y.~Takeuchi}

\author[KRISS]{Y.~H.~Kim}
\author[KRISS]{K.~B.~Lee}
\author[KRISS]{M.~K.~Lee}
\author[KRISS]{J.~S.~Lee}

\author[MIYA]{Y.~Fukuda}

\author[NAGOYA,KMS]{Y.~Itow}
\author[NAGOYA]{Y.~Nishitani}
\author[NAGOYA]{K.~Masuda}
\author[NAGOYA]{H.~Takiya}
\author[NAGOYA]{H.~Uchida}

\author[SEJONG]{N.~Y.~Kim}
\author[SEJONG]{Y.~D.~Kim}

\author[TOKAI1]{F.~Kusaba}
\author[TOKAI2]{D.~Motoki\fnref{tohoku}}
\author[TOKAI1]{K.~Nishijima}

\author[YNU]{K.~Fujii}
\author[YNU]{I.~Murayama}
\author[YNU]{S.~Nakamura}

\address[ICRR]{Kamioka Observatory, Institute for Cosmic Ray Research,
  the University of Tokyo, Higashi-Mozumi, Kamioka, Hida, Gifu, 506-1205, Japan}
\address[GIFU]{Information and Multimedia Center, Gifu University, Gifu 501-1193, Japan}
\address[IPMU]{Kavli Institute for the Physics and Mathematics of the Universe,
  the University of Tokyo, Kashiwa, Chiba, 277-8582, Japan}
\address[KMS]{Kobayashi-Maskawa Institute for the Origin of Particles and the Universe, 
Nagoya University, Furo-cho, Chikusa-ku, Nagoya, Aichi, 464-8602, Japan.}
\address[KOBE]{Department of Physics, Kobe University, Kobe, Hyogo 657-8501, Japan}
\address[KRISS]{Korea Research Institute of Standards and Science, Daejeon 305-340, South Korea}
\address[MIYA]{Department of Physics, Miyagi University of Education, Sendai, Miyagi 980-0845, Japan}
\address[NAGOYA]{Solar Terrestrial Environment Laboratory, Nagoya University, 
Nagoya, Aichi 464-8602, Japan}
\address[SEJONG]{Department of Physics, Sejong University, Seoul 143-747, South Korea}
\address[TOKAI1]{Department of Physics, Tokai University, Hiratsuka,
  Kanagawa 259-1292, Japan}
\address[TOKAI2]{School of Science and Technology, Tokai University, Hiratsuka,
  Kanagawa 259-1292, Japan}
\address[YNU]{Department of Physics, Faculty of Engineering, Yokohama National University, Yokohama, Kanagawa 240-8501, Japan}

\fntext[tohoku]{Now at Research Center for Neutrino Science, Tohoku University, Sendai 980-8578, Japan}